\documentclass[conference]{IEEEtran}

\IEEEoverridecommandlockouts
\usepackage{amssymb}
\usepackage{booktabs}
\usepackage{verbatim}
\usepackage{cite}
\usepackage{amsmath}
\usepackage{tikz}
\usetikzlibrary{matrix}
\usepackage{threeparttable}
\usepackage{footnote}
\usepackage{subfigure}
\usepackage{pgfplots} % package used to implement the plot
\pgfplotsset{width=8cm, compat=1.6}
\usepgfplotslibrary{groupplots}
\usetikzlibrary{matrix}
\usepgfplotslibrary{external}

\usepackage{url}
\begin{document}

%\title{Link Prediction Based on User Information and Topology in Social Networks}
\title{LP-UIT: A Multimodal Framework for Link Prediction in Social Networks}

\begin{comment}
\author{\IEEEauthorblockN{Huizi Wu}
\IEEEauthorblockA{School of Information Management\\ and Engineering\\
Shanghai University of Finance \\and Economics, China\\
wuhuizisufe@gmail.com}
\and
\IEEEauthorblockN{Shiyi Wang}
\IEEEauthorblockA{School of Information Management\\ and Engineering\\
Shanghai University of Finance \\and Economics, China\\
wang\_shiyi1998@163.com}
\and
\IEEEauthorblockN{Hui Fang$^*$}
\IEEEauthorblockA{School of Information Management\\ and Engineering\\
Shanghai University of Finance \\and Economics, China\\
fang.hui@mail.shufe.edu.cn}

\thanks{$^*$Corresponding author.}
}
\end{comment}

\author{\IEEEauthorblockN{Huizi Wu\IEEEauthorrefmark{1}, Shiyi Wang\IEEEauthorrefmark{2}, and Hui Fang\IEEEauthorrefmark{3}}
\IEEEauthorblockA{
Research Institute for Interdisciplinary Sciences, 
School of Information Management and Engineering\\
Shanghai University of Finance and Economics, China\\
\IEEEauthorrefmark{1}wuhuizisufe@gmail.com,
\IEEEauthorrefmark{2}wang\_shiyi1998@163.com,
\IEEEauthorrefmark{3}fang.hui@mail.shufe.edu.cn}

\thanks{\IEEEauthorrefmark{3}Hui Fang is corresponding author.}
}


%Telephone: (800) 555--1212, Fax: (888) 555--1212}


%\IEEEkeywords{social network}

\maketitle

\begin{abstract}
With the rapid information explosion on online social network sites (SNSs), it becomes difficult for users to seek new friends or broaden their social networks in an efficient way. Link prediction, which can effectively conquer this problem, has thus attracted wide attention. Previous methods on link prediction fail to comprehensively capture the factors leading to new link formation: 1) few models have considered the varied impacts of users' short-term and long-term interests on link prediction. Besides, they fail to jointly model the influence from social influence and ``weak links''; 2) considering that different factors should be derived from information sources of different modalities, there is a lack of effective multi-modal framework for link prediction. In this view, we propose a novel multi-modal framework for link prediction (referred as LP-UIT) which fuses a comprehensive set of features (i.e., user information and topological features) extracted from multi-modal information (i.e., textual information, graph information, and numerical information). Specifically, we adopt graph convolutional network to process the network information to capture topological features, employ natural language processing techniques (i.e., TF-IDF and word2Vec) to model users' short-term and long-term interests, and identify social influence and ``weak links" from numerical features. We further use an attention mechanism to model the relationship between textual and topological features. Finally, a multi-layer perceptron (MLP) is designed to combine the representations from three modalities for link prediction. Extensive experiments on two real-world datasets demonstrate the superiority of LP-UIT over the state-of-the-art methods.

\end{abstract}

\begin{IEEEkeywords}
link prediction, social network, multi-modal framework, social influence
\end{IEEEkeywords}

% \fi
%\IEEEpeerreviewmaketitle

\section{Introduction}
With the development of information technology, people can no longer live without the Internet and feel increasingly comfortable to interact on online social networks with varied activities, such as following other people, paying attention to people/things they are interested in, giving public opinions, and building their virtual social networks. All of these behaviors lead to a huge amount of data produced in every minute \cite{he2017latent}. While it brings about enormous research opportunities, the massive amount of data also incurs lots of difficulties in analyzing and exploring social network sites (SNSs), among which the most challenging issue is the information overload problem, which is considered as a hot topic for both academic and business applications. 

%In order to achieve goals like expanding users' social network, we need to filter data to get the really useful and important information and ignore those we do not care about. For this reason, how to solve the problem of information overloading is considered as hot topics for both academics and business.
Link prediction, as one of the effective methods towards tackling the information overload problem on social network sites,
%The main method to solve this problem is link prediction, which
aims to predict the likelihood of a possible directed link between two users in the future. For practical applications, on the one hand, link prediction in a social network can help a user to expand his/her social network and thus improve his/her loyalty on the corresponding social network site. On the other hand, it can also be used to target advertisements and bring more profits for the platform. Consequently, almost all of the existing social network sites nowadays have their own systems for better link prediction and 
social recommendation. For instance, QQ\footnote{\url{im.qq.com}.}, Facebook\footnote{\url{facebook.com}.} and LinkedIn\footnote{\url{www.linkedin.com}.} all have functions that can recommend ``people you may know" to users.

There are different social networks in general. For example, Katz et al.\cite{katz1973uses} stated that users involve in social media with different attributes to satisfy needs of five general categories: information, emotion, connection, integrative, and escape. Mikyeung \cite{bae2018understanding} also divided users' basic motivations on using SNSs into six categories: seeking information, seeking entertainment, convenience, seeking socialization, seeking social support, and escapism. In this study, we mainly focus on the link prediction problem on social network sites which can satisfy users’ \emph{information seeking needs}. 

In information-seeking oriented social networks, there are two possible motivational factors (i.e., individual level and peer level) that drive each user to get connected with other users. That is, for a user, on the individual level, he/she may want to get information, learn knowledge, or improve himself/herself with intrinsic motivation. For this purpose, he/she is mainly concerned about other users' performance and competence (capability), which might be measured by ones' activities in the social network such as public opinions on other items. We conclude this as \emph{user information}.
On the peer level, the user can judge other users from their linked peers (friends), including the common friends with other users, the shortest path, and so on. We conclude this factor as \emph{topological information} in our study.
Therefore, we consider that, in such kinds of social networks (for the purpose of information-seeking), link prediction problems should incorporate both \emph{user information} and \emph{topological information}.

%In peer level, users want to find some information from their friends and environments. The peer level can be defined as user's social structures such as the shortest path, common friend, the interaction between users etc., which are referred to topological features. Therefore, for any information seeking social networks, link prediction problem can be solved based on user information and their topological information.

Towards the specific link prediction problem, previous studies mainly suffer from the following limitations: 1) quite a few factors regarding user information and topological information have been ignored by previous methods. For example, few existing methods have taken users' possibly changing interests into consideration. That is, they ignored that users' preferences/interests on different topics are dynamic and might evolve over time. Besides, few of them have considered the impact of weak links (instead of direct links) between users in peer level; 2) there is a lack of effective methods to combine different features on the two perspectives (i.e., peer level and individual level), which are mainly derived from different source information of different modality. Although some recent deep learning methods (e.g., TADW \cite{yang2015network} and graph neural network based methods \cite{liao2018attributed, pan2018adversarially, simonovsky2018graphvae, ma2018constrained, de2018molgan}) have been proposed for link prediction, they either focus on one modality of information (e.g., text information in TADW \cite{yang2015network}), or cannot easily and effectively fuse user information \cite{liao2018attributed, de2018molgan}.

%However, there are still many problems in link prediction. Firstly, there are many factors that are not considered in link prediction. For example, few of existing methods have taken account of the changing interests of users which means few of them classify users’ interests by time in individual level. Besides, in individual level, only a few researchers take user social influence into consideration. In addition, existing link prediction methods also do not consider the impact of weak links between users in peer level. Secondly, there is no very good 
%method for link prediction to combine multiple information. TADW \cite{yang2015network} proposes a framework combine text information into the matrix factorization. This method only suit text information rather than multimodal data. Furthermore, most graph neural network methods \cite{liao2018attributed, pan2018adversarially, simonovsky2018graphvae, ma2018constrained, de2018molgan} treat user information as the initial embedding vector, which weaken the role of domain features.

%In this paper, we propose a method to think over all of these factors. 
Therefore, in view of the aforementioned issues, we propose a novel end-to-end learning method called \underline{L}ink \underline{P}rediction based on \underline{U}ser \underline{I}nformation and \underline{T}opology (referred as LP-UIT). Specifically, we first learn short-term and long-term interests for each user from textual information. We then design an end-to-end multi-modal framework for link prediction, which considers all factors on individual level and peer level derived from different information sources of different modality, including users' interests (textual information), topological features (graph information), social influence and weak links (numerical information). In particular, we adopt graph convolutional network (GCN) to represent the topological information, and Word2Vec to represent users' interests from textual information. We further model the relationship between textual information and graph information using attention mechanism. Finally, we use a multi-layer perceptron (MLP) to fuse the three 
types of features for final link prediction. %More importantly, we propose to use the attention mechanism to obtain better performance.
%Then we propose that users’ interaction (weak links) will have impacts on forming links. All of these factors are considered in our framework.

The main contributions of this work are three-fold:
\begin{itemize}
\item To capture users' dynamic interests, we take both the short-term and long-term interests into consideration and measure their different impacts on link prediction problem. Besides, we consider the impact of both social influence and weak links. 

\item We design a novel multi-modal model LP-UIT which fuses multiple factors extracted from different information sources of three modalities, including the short-term and long-term interests (textual information), users’ social influence and weak links (numerical information), and network structure information (graph information). We further design an end-to-end learning framework for link prediction.

\item Extensive experiments on two real-world datasets show that our model outperforms state-of-the-art methods.
\end{itemize}

%%%%%%%%%%%%%%%%%%%%%%%%%%%%%%%%%%%%%%%%%%%%%%%%%%%%%%%%%%%%%%%%%%%%%%
\section{Related Work}
%%%%%%%%%%%%%%%%%%%%%%%%%%%%%%%%%%%%%%%%%%%%%%%%%%%%%%%%%%%%%%%%%%%%%%
Previous studies on link prediction can be divided into four main categories: similarity-based methods, probabilistic methods, relational models, and learning-based methods \cite{mutlu2019review}.
Our study is mainly related to similarity-based methods and learning-based methods.

\subsection{Similarity-based methods}

\emph{Similarity-based methods} (i.e., proximity-based methods) can be grouped into nodal proximity-based methods and structural proximity-based methods. 

For nodal proximity-based methods, the basic assumption is that if users have more similar interests, they are more likely to form a link in the future. There are multiple ways to measure users' interests, while several studies proposed to measure user similarity in terms of users' long-term and short-term interests separately. For example, Li et al. \cite{li2014modeling} stated that users would have a relatively stable preference in long-term, whereas short-term preference would change over time.
%Using this theory to recommend a new item can have a more promising performance than other existing methods. 
Yin et al. \cite{yin2014temporal} observed that users’ behaviors are influenced by users’ intrinsic interests (identical to long-term) and public attention (short-term), since users’ intrinsic interests are relatively stable while the attention of the public changes. 
%They use this theory to model users’ behaviors and prove it indeed performs better. 
Other studies \cite{spasojevic2014lasta, jiang2015modeling, piao2018inferring} also found out that users have long-term and short-term interests while the long-term interests are always related to users’ intrinsic attributes and the short-term interests are connected with the hot topics or events at the moment. Besides, Wellman \cite{wellman1997electronic} stated that lots of time and mutual investment can build strong social relationships. Thus, people are more likely to share information, reviews, and decisions with people who have strong relationships with them. That is, users are more likely to build links with those who have more interactions with them. Zhang et al. \cite{zhang2011intrank} designed various interaction attributes to recommend friends and proved that the accordingly proposed approach had an advantage over traditional similarity-based methods. 

For structural proximity-based methods, they considered that users' social influence in social networks has a huge impact on link prediction. For example, Kelman \cite{kelman1958compliance} identified social influence as compliance, identification and internalization. Marsden et al. \cite{marsden1993network} suggested that social influence can alter users’ responses, and the proximity of users can also be affected by interpersonal influence between users. Based on these theories, Huo et al. \cite{huo2018link} advanced the link prediction method with social influence.
%and show the method could perform better.

Although different factors have been considered by different studies as discussed, there are few studies that have simultaneously considered and distinguished both long-term and short-term interests, whilst jointly considered both strong relationship and ``weak ties". That is to say, some factors regarding user information and topology information have been dismissed in previous studies. On the other hand, other than the network (graph) information, most of the activities in information-seeking oriented social networks are presented in other modalities (e.g., textual information and numerical information), which are greatly ignored in previous studies. 
Therefore, towards a comprehensive framework on link prediction, we propose a multi-modal model to jointly consider factors extracted from different modalities.

\subsection{Learning-based methods}

Previous \emph{learning-based methods} can be concluded into classification-based (e.g., SVM and decision tree), matrix factorization-based, random walk-based (DeepWalk \cite{perozzi2014deepwalk}, Node2Vec \cite{grover2016node2vec}, LINE \cite{tang2015line}), and graph neural network-based (GNN) methods. In particular, GNN-based methods have adopted graph convolutional networks \cite{kipf2017semi}, graph gated neural networks \cite{li2016gated}, hierarchical graph embedding methods \cite{ying2018hierarchical} and graph attention networks \cite{velickovic2018graph}, respectively. 

As for the neural network-based methods (including GNN-based), they can be categorized into: methods based only on network structure and methods that combine multiple information. Earlier methods (MDS \cite{cox2008multidimensional}, LLE \cite{roweis2000nonlinear}, and ISOMAP \cite{tenenbaum2000global}) are shallow models which extract node representations from an affinity graph. For better node representation, DeepWalk uses local information obtained from truncated random walks to learn latent representations. Node2Vec (based on the DeepWalk) learns  mappings of nodes as a low-dimensional space of features that maximizes the likelihood of preserving network neighborhoods of nodes by a random walk. LINE defines two kinds of similarities on the graph (i.e., first-order similarity and second-order similarity) to obtain node representations. Although these methods have made some progress, they are difficult to capture the non-linear structure of the network. Subsequently, many deep models \cite{cao2016deep,wang2016structural,liao2018attributed} 
including GNNs have been proposed for network embedding. However, most GNN-based methods treat user information as an embedding vector in the input layer, but ignore the role of user information. They also fail to effectively consider textual information and numerical features, e.g., social influence and interactions between every corresponding user pair. 

\begin{figure*}[htbp]
    \centering
    \includegraphics[width=18.5cm]{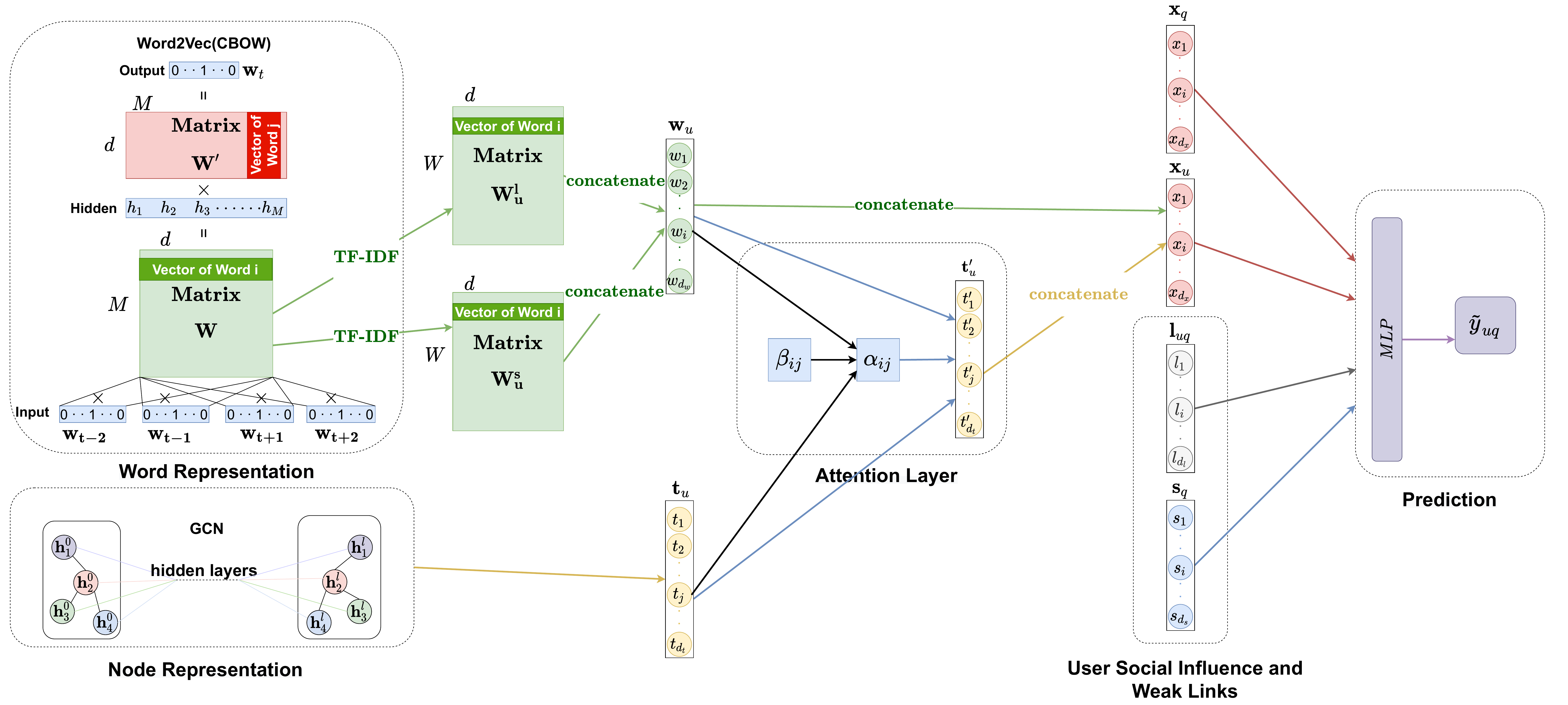}
    \caption{The overview of our proposed LP-UIT model.}
    \label{fig:model}
\end{figure*}

On the other side, some methods connect the text with the adjacent vector and then use a joint vector for classification. For example, TADW \cite{yang2015network} proposes a framework that combines text information into matrix factorization. However, this method only suits text information rather than multi-modal data \cite{huang2018multimodal}. Furthermore, TADW cannot handle large network structure due to the high computational complexity. Recently, there are some multi-modal methods to combine link information with information of other modality. For example, AMVAE \cite{huang2019network} fuses the links and multi-modal contents for network embedding. However, this method aims to classify nodes (i.e., determine labels on pictures) instead of solving the link prediction problem. Similarly, deepMDBN \cite{huang2018multimodal} extracts topological features and word vectors, then proposes a three-layer neural network for link prediction. However, it ignores to incorporate the numerical features. To the best of our knowledge, there is no specific multi-modal framework (that simultaneously considers three modalities) for link prediction problem in social networks.

%%%%%%%%%%%%%%%%%%%%%%%%%%%%%%%%%%%%%%%%%%%%%%%%%%%%%%%%%%%%%%%%%%%%%%
\section{The LP-UIT Model}

In this section, we will first define our research problem and then present the LP-UIT model detailedly.

%%%%%%%%%%%%%%%%%%%%%%%%%%%%%%%%%%%%%%%%%%%%%%%%%%%%%%%%%%%%%%%%%%%%%%
\subsection{Problem Formulation}
We treat a social network at the particular time $t$ as a directed graph $\mathcal{G}_t = (\mathcal{V}_t,\mathcal{E}_t)$, where $\mathcal{V}_t=\{v_1,v_2,\cdots, v_N\}$ denotes the set of nodes ($|\mathcal{V}|=N$), i.e. all users in this social network and $\mathcal{E}_t$ is the set of all edges, i.e. all directed links between users.

The link prediction aims to predict the probability of any new links between users in a future time $t’$, where $t’ > t$. Then, we recommend a set of new links (i.e., $K$) with the highest probability values in terms of social network $\mathcal{G}_t$.

\subsection{Our Model}
Here, we describe our multi-modal framework in detail. As shown in Figure \ref{fig:model}, we firstly use Word2Vec \cite{mikolov2013efficient} to obtain word representation (long-term and short-term interests). We also process data to get user social influence and weak links between users. Secondly, we use the graph convolutional network (GCN) model to learn node embedding. Thirdly, we use an attention mechanism to fuse word embedding with node embedding, and 
identify the relationship between them. After obtaining the representation of each user, we use a multi-layer perceptron (MLP) to combine weak links and social influence with users' embeddings for link prediction.

In the following subsections, we will introduce each component.

\subsubsection{Word Representation}
%After collecting all data we need, we begin to process data and get users interests, their social influence, weak links between them and the exited links. 

%%%%%%%%%%%%%%%%%%%%%%%%%%%%%%%%%%%%%%%%%%%%%%%%%%%%%%%%%%%%%%%%%%%%%%
%\subsubsection{Long-term and Short-term Interests}
As we mentioned before, users have two types of interests: one is relatively stable, and the other changes from time to time depending on the attention of the public. We call these two interests long-term interests and short-term interests respectively. Specifically, the long-term interests of a user will be extracted from his/her all past activities happened during $[0,t]$, whilst the short-term interests are derived from his/her recent activities happened during $[t_r,t]\ (t-t_r\ll t)$.

For every user $u$, we use TF-IDF \cite{salton1975vector} method to extract $W$
%\hui{notations could be modified.} 
textual words from the corresponding activities with the highest $TF-IDF$ values as his/her long-term interests, and extract the same number of words to represent his/her short-term interests accordingly. 
%\hui{What are the specific textual words for long-term and short-term preferences respectively?}  
%The number of words varies with different datasets, Zhihu dataset has 5 topics --- user answers, columns user followed, questions users asked and so on, and we extract 10 words for every topic. For Epinions dataset, we extract 20 words.

The idea of TF-IDF is presented as follows: assuming $M$ is the number of all documents in corpus, $m(i)$ is the number of documents that have word $i$, and $\text{frequency}(i,j)$ is the frequency for word $i$ in documents $j$; $k$ refers to any of the words in document $j$. Thus, $TF$, $IDF$, and $TF-IDF$ are calculated as Equations~\ref{tf},~\ref{idf} and~\ref{tf-idf}, respectively.
\begin{equation}
TF(i, j)=\frac{\text{frequency}(i, j)}{\sum_{k} \text {frequency}(k, j)}
\label{tf}
\end{equation}
\begin{equation}
IDF(i)=\log \frac{M}{m(i)}
\label{idf}
\end{equation}
\begin{equation}
TF-IDF(i,j)= TF(i,j) \times IDF(i)
\label{tf-idf}
\end{equation}

We then use Word2Vec to map these words into continuous vectors (each word representation $\in \mathbb{R}^{d}$) and concatenate these vectors to form two representations, which imply user $u$'s long-term interests $\mathbf{w}_u^l$ and short-term interests $\mathbf{w}_u^s$, respectively. Finally, we concatenate these two representations of user $u$ to obtain his/her overall word representation, i.e., $\mathbf{w}_u\in \mathbb{R}^{d_w}=concate{(\mathbf{w}_u^s,\mathbf{w}_u^l)}$, where $d_w=2W\times d$.

%%%%%%%%%%%%%%%%%%%%%%%%%%%%%%%%%%%%%%%%%%%%%%%%%%%%%%%%%%%%%%%%%%%%%%
\subsubsection{User Social Influence and Weak Links (Numeric Feature Representation)}

As aforementioned, user $u$’s behaviors and attitudes could be influenced by other users. In other words, social influence $\mathbf{s}_u\in\mathbb{R}^{d_s}$ exerts an impact on new link formation.
%Hence, we consider users’ social influence when predicting links. 
In an online social network, we consider social influence being identical to agreeing to content other people published, voting for other people, giving other people comments, concerning other people’s questions, etc.

%%%%%%%%%%%%%%%%%%%%%%%%%%%%%%%%%%%%%%%%%%%%%%%%%%%%%%%%%%%%%%%%%%%%%%
%\subsubsection{Weak Link}
On the other hand, user $u$ also has ``weak links” with another user $q$ (denoted as $\mathbf{l}_{uq}\in\mathbb{R}^{d_l}$) even there is no real link between the two users at time $t$. That is to say, in the absence of directed links, they may still have interactions with each other such as user $u$ may answer questions asked by user $q$. And we argue that such kind of weak links can also affect the formation of new real links \cite{zhang2011intrank}. We further classify weak links into two factors: the \emph{quantity} of weak links and the \emph{quality} of weak links. Specifically, the quantity of interactions is the number of such kind of interactions between two corresponding users, e.g., how many times user $u$ answered user $q$'s questions. On the contrary, the quality of weak links indicates the competence of user $q$, e.g., how many likes user $q$ got from all of these answers.  

%%%%%%%%%%%%%%%%%%%%%%%%%%%%%%%%%%%%%%%%%%%%%%%%%%%%%%%%%%%%%%%%%%%%%%
\subsubsection{Node Representation}
We use graph convolutional network (GCN) to obtain node representation. Specifically, we use a linear layer to train nodes' one-hot vectors and get $\mathbf{H}^0 \in \mathbb{R}^{N\times d_{t0}}$ as our initial embedding matrix of node set (initial node embedding $\mathbf{h}^0_u=\mathbf{H}^0_{u,:}\in\mathbb{R}^{d_{t0}}$).

%We use classical graph convolutional network to obtain node representation. Specifically, we denote $\mathbf{H}^0 \in \mathbb{R}^{N\times d_{t0}}$ as initial embedding matrix of item set (initial node embedding $\mathbf{h_u}^0=\mathbf{H}^0_{u,:}\in\mathbb{R}^{d_{t0}}$).

\begin{comment}
\begin{equation}
\small
\mathbf{H}^0=\text{nn.Embedding}(N, d_{t0})\label{eq:initialnnEmbedding}\vspace{-1mm}
\end{equation}
\end{comment}

Given graph $\mathcal{G}$, we can obtain adjacency matrix $\mathbf{A}\in\mathbb{R}^{N\times N}$ and degree matrix $\mathbf{D}\in\mathbb{R}^{N\times N}$ (a diagonal matrix whose elements are degrees for corresponding nodes). Then, the layer-wise propagation rule is as follows:
\begin{equation}
\mathbf{H}^{l+1}=\phi (\tilde{\mathbf{D}}^{-\frac{1}{2}} \tilde{\mathbf{A}} \tilde{\mathbf{D}}^{-\frac{1}{2}} \mathbf{H}^l \mathbf{W}^l) \label{GNN}
\end{equation}
where $\tilde{\mathbf{A}}= \mathbf{A} + \mathbf{I}$,
%$\tilde{\mathbf{A}}$ equals the adjacency matrix $\mathbf{A}$ plus the identity matrix $\mathbf{I}$,
i.e. $ \tilde{\mathbf{A}}$ is $\mathbf{A}$ plus self-connections. $ \tilde{\mathbf{D}}\in\mathbb{R}^{N\times N}$ is the degree matrix of $\tilde{\mathbf{A}}$, $\mathbf{W}^l\in\mathbb{R}^{d_{tl}\times d_{tl}}$ is a layer-specific trainable weight matrix. $\phi$ denotes an activation function, such as ReLU function. $\mathbf{H}^l\in\mathbb{R}^{N\times d_{tl}}$ is the matrix of node representation in the l-th layer. Then the final node embedding $\mathbf{t}_u= \mathbf{H}_{u,:}$ and $t_u\in\mathbb{R}^{d_t}$
%%%%%%%%%%%%%%%%%%%%%%%%%%%%%%%%%%%%%%%%%%%%%%%%%%%%%%%%%%%%%%%%%%%%%%
\subsubsection{Attention Layer}
After Word2Vec and a two-layer GCN, for each user $u$, we can obtain the corresponding word representation $\mathbf{w}_u\in \mathbb{R}^{d_w}$, where $\mathbf{w}_u=\{w_1,...,w_i, ...w_{d_w} \}$ and node representation $\mathbf{t}_u  \in \mathbb{R}^{d_t}$, where $\mathbf{t}_u=\{t_1,...,t_j, ...t_{d_t} \}$. %Then we concate the short-term interest and long-term interest to form word representation $\mathbf{w}_u\in \mathbb{R}^{d_w}=concate{(\mathbf{w}_u^s,\mathbf{w}_u^l)}$, where $\mathbf{w}_u=\{w_1,...,w_i, ...w_{d_w} \}$.
Then, we design an attention layer to obtain his/her final representation, which aims to automatically find the correlation between user and topology information.
After getting embeddings, the self-attention coefficient $e_{ij}$ between each $w_i$ and topology $t_j$ is computed as:
\begin{equation}
e_{ij}=\phi (w_i \beta_{ij} t_j) \label{self-attention}
\end{equation}
where $\phi(.)$ is the ReLu function and $\beta_{ij}$ is the parameter to be learned. The softmax function is further adopted to normalize $e_{ij}$:
\begin{equation}\small
\alpha_{ij}=\text{softmax}(e_{ij})=\frac {exp(e_{ij})}{\sum\limits_{k} exp(e_{kj})}\label{eq:normalizedAttentionValue}%\vspace{-1mm}
\end{equation}

The weight $\alpha_{ij}$ measures the impact of the user feature $w_i$ on topology feature $t_j$. Then, the weighted node embedding $\mathbf{t}_u^{\prime}$ can be represented as:
\begin{equation}\small
    \mathbf{t}_u^{\prime}=\{t_1^{\prime},...,t_j^{\prime}, ...t_{d_t}^{\prime} \}
\label{tu}%\vspace{-1mm}
\end{equation}
where
\begin{equation}\small
t_j^{\prime}= \sum\limits_{i} \alpha_{ij} w_{i} t_j
\label{eq:normalizedAttentionValue}%\vspace{-1mm}
\end{equation}

Finally, we concatenate word embedding and weighted node embedding to form the final user embedding $\mathbf{x}_u$. 
\begin{equation}\small
\mathbf{x}_u= concate(\mathbf{w}_u, \mathbf{t}_u^{\prime})\label{eq:userEmbedding}%\vspace{-1mm}
\end{equation}

%%%%%%%%%%%%%%%%%%%%%%%%%%%%%%%%%%%%%%%%%%%%%%%%%%%%%%%%%%%%%%%%%%%%%%
\subsubsection{Prediction}
In the Attention Layer, we obtain the embedding $\mathbf{x}$ for each user. Then, we adopt a MLP to get the link probability from user $u$ $\to$ user $q$, whose inputs are user $u$'s embedding $\mathbf{x}_u$, user $q$'s embedding $\mathbf{x}_q$, user $q$'s social influence $\mathbf{s}_q$, and weak links between $u$ and $q$ $\mathbf{l}_{uq}$:
\begin{equation}\small
\tilde{y}_{uq}=MLP(\mathbf{x}_u,\mathbf{x}_q,\mathbf{s}_q,\mathbf{l}_{uq})\label{eq:LinkProbability}%\vspace{-1mm}
\end{equation}

We adopt cross-entropy loss to train the model:
\begin{equation}\small
L=-\sum y_{uq} \log(\tilde{y}_{uq})+(1-y_{uq})\log (1-\tilde{y}_{uq}) \label{eq:loss}\vspace{-1mm}
\end{equation}
where $y_{uq}$ is the ground-truth of u $\to$ q.

\section{Experiments}
\label{sect_figure}
In this section, we conduct experiments on two datasets to validate the effectiveness of our model by answering the following research questions (RQs):

\begin{itemize}
\item \textbf{RQ1}: How does LP-UIT perform compared to other state-of-the-art methods?
%\item \textbf{RQ2}: Is it useful to incorporate short term interest and long term interest to our model? 
% \item \textbf{RQ3}: How does attention layer of LP-UIT contribute to the performance? 
\item \textbf{RQ2}: How do different components of LP-UIT (e.g., short-term or long-term interests) contribute to the performance? 
\item \textbf{RQ3}: How do different hyper-parameters affect the performance of LP-UIT?
\end{itemize}

%We first describe the experimental settings. Then we answer the above research questions one by one.

\subsection{Experimental Setup}

\subsubsection{Datasets}

We use two datasets in our experiment: Zhihu dataset
%,Twitter-Dynamic-Action in AMiner datasets\footnote{\url{https://www.aminer.cn/data-sna}.} 
and Epinions dataset\footnote{It was provided by Dr. Jiliang Tang and previously publicized on his website (\url{www.cse.msu.edu/~tangjili}).}. The details of the two datasets are shown in Table~\ref{table1}.

\begin{itemize}
    \item[$\bullet$]
    \textbf{Zhihu:} Zhihu\footnote{\url{www.zhihu.com}.} is a social network site where users can ask and answer questions of their interests. Users can also follow or be followed by other people in Zhihu. We totally crawl $11,983$ unique users with corresponding personal information. After filtering out some inactive users, we finally get $11,114$ unique users.
    
    \item[$\bullet$]
    \textbf{Epinions:} Epinions is a website where users can read other customers' ratings and reviews. Users can also choose whether to trust other users or not.
    %In the meantime, they can also be trusted by other customers. 
\end{itemize}

\begin{table}[htb]
\normalsize
\centering
\caption{Statistics of Experimental Datasets}
\label{table1}
\begin{tabular}{ccc}
\toprule

\textbf{Datasets} & \textbf{Zhihu} & \textbf{Epinions}\\
\hline

\# of Users & 11,114 & 12,772\\
\# of Links at time $t$ & 252,967 & 240,585\\
%\# of Links at time $t'$ & 54160 & 60146\\%
%Average Degree & 27.63 & 23.54\\%
\# of Links at time $t'$ & 307,127 & 300,731\\

\bottomrule
\end{tabular}
%}%\vspace{-3mm}
\end{table}

For two datasets, we extract four types of features as model input: long-term and short-term interests, user social influence, weak links, and structure information. To classify users' interests into long-term and short-term, we use each user’s most recent activities (the latest top $10\%$ data) to represent his/her short-term interests $\mathbf{w}^s$, and all of his/her activities to represent their long-term interests $\mathbf{w}^l$. Furthermore, the number of words (i.e., $W$) varies across different datasets. Since Zhihu dataset has $5$ topics (user answers, columns user followed, questions users asked, topics user followed, and questions user followed), we extract $10$ words for every topic (i.e., $W=50$). For Epinions dataset, we extract $20$ words with the highest $TF-IDF$ ($W=20$). 

After extracting words, we found that, on Zhihu, the typical words on long-term interests are topics related to users' personal preferences, such as 2-D world, Lumbar disc herniation, Financing, Gourmet food, and so on. In contrast, the typical words on short-term interests change over time and are more related to hot and public topics that happened recently, like \textit{Donald Trump}, blind date, unemployed, freshman, and \textit{Teacher's Day}.

\subsubsection{Baseline Methods}
To demonstrate the effectiveness of our model, we compare it with the following state-of-art approaches:
\begin{itemize}
    \item[$\bullet$]
    \textbf{CN (Common Neighbors)} \cite{newman2001clustering}: it is a basic graph analysis algorithm, and obtains the common neighbors between two nodes to rank the possible new links.

    \item [$\bullet$]
    \textbf{PR (PageRank)} \cite{tong2006fast}: the basic idea of PageRank algorithm is to define a random walk model (the first-order Markov chain) on a directed graph, which describes the behavior of random walkers randomly visiting each node along with the directed graph. The value indicates the importance of the node.

     \item [$\bullet$]
     \textbf{Node2vec} \cite{grover2016node2vec}: it is a biased random walk algorithm based on DeepWalk \cite{perozzi2014deepwalk} for graph embedding, which applies two hyper-parameters $p$ and $q$ to control the random walk. Note that when $p$ and $q$ are set to $1$, node2vec equals to DeepWalk.
     
     \item [$\bullet$]
     \textbf{TADW} \cite{yang2015network}: it adopts matrix factorization to incorporate rich text information into the network embedding.
     
     \item [$\bullet$]
     \textbf{DeepMDBN} \cite{liu2015multimodal}: it is a multi-modal framework and predicts link values by jointly considering textual information (e.g., user comments) and network structure (e.g., in-degree and out-degree).

     \item [$\bullet$] 
     \textbf{ARGA} \cite{pan2018adversarially}: it builds a novel adversarial graph embedding framework in a graph by encoding the topological structure and node content to a compact representation, on which the decoder is trained to reconstruct the graph structure.      
     
\end{itemize}

%%%%%%%%%%%%%%%%%%%%%%%%%%%%%%%%%%%%%%%%%%%%%%%%%%%%%%%%%%%%%%%%%%%%%%
\begin{table*}[htb]
\normalsize
\centering
\caption{Performance of all methods on the two datasets. The best performance is boldfaced, and the runner up is underlined. We compute the improvements that LP-UIT achieves relative to the best baseline. Statistical significance of pairwise differences of LP-UIT vs. the best baseline is determined by a paired t-test ($^{***}$ for p-value $\leq$.01,$^{**}$ for p-value $\leq$.05,$^{*}$ for p-value $\leq$.1).}\label{tb:comparativeResults}%\vspace{-3mm}
\begin{tabular}{cc|cccc|cccc }
\hline

&&\multicolumn{4}{c}{Zhihu}&\multicolumn{4}{|c}{Epinions}\\

\hline

& \textbf{Methods}  & \textbf{AUC} & \textbf{  KS} & \textbf{NDCG@500} & \textbf{MAP@500} & \textbf{AUC} & \textbf{    KS} & \textbf{NDCG@500} & \textbf{MAP@500} \\
\hline

&CN       &0.6870  &0.3741  &0.6798  &0.6860  &0.8119  &0.6320  &0.7203  &0.7120 \\

&PR        &0.5974  &0.1960  &0.5538  &0.5700  &0.5863  &0.1745  &0.8640  &0.8564\\

&node2vec  &0.6705  &0.3624  &0.1669  &0.1659  &0.5577  &0.085  &0.1246  &0.1221 \\

&TADW      &0.5289  &0.1395  &0.6232  &0.7888  &0.7764  &0.4193  &0.8390  &0.8642 \\

&deepMDBN  &\underline{0.9369}  &\underline{0.7755}  &0.8864  &0.8597  &0.8538  &0.5564  &\underline{0.9396}  &\underline{0.9486} \\

&ARGA     &0.8532  &0.6217  &\underline{0.9155}  &\underline{0.9130}  &\underline{0.8940}  &\underline{0.6474}  &0.8922  &0.8629 \\

\hline
&LP-UIT    &\textbf{0.9516}  &\textbf{0.7815}  &\textbf{0.9358}  &\textbf{0.9212}  &\textbf{0.9048}  &\textbf{0.6594}  &\textbf{0.9750}  &\textbf{0.9772}
\\
\hline
&Improvement & 1.57\%*** & 0.77\% & 2.22\%** & 0.90\% & 1.21\%*** & 1.85\%*** & 3.77\%*** & 3.01\%*** \\

\hline
\end{tabular}
%}%\vspace{-3mm}
\end{table*}

\subsubsection{Evaluation Metrics}
We compare the performance of different methods on link prediction in terms of the following four accuracy-related metrics. Noted that for the four metrics, larger values indicate better performance:

\begin{itemize}
    \item[$\bullet$]
    \textbf{AUC (Area under the ROC Curve)}: AUC is to measure the prediction accuracy as a whole. The idea is to randomly select edges $n$ times from the set of non-existent edges and test set respectively and compare the predicted scores that a model generates toward each corresponding edge pair. Let $n'$ represents the number of times that edge score in the test set is larger than that in non-existent edge set and $n''$ represents the number of times that the two edge scores are equal.
    %The formula of AUC is Equation~\ref{AUC}. For AUC method, if AUC score is greater than 0.5 and is closer to 1,the method is better. When AUC score is equal to 0.5, the method using to predict links is meaningless.
 \begin{equation}
AUC= \frac{n'+0.5n''}{n}
\label{AUC}
\end{equation}   
%    \item [$\bullet$]
%    \textbf{F1:} F1 indicator can be regarded as the harmonic average of accuracy and recall of the model. The equation of F1 shows in Equation~\ref{F1}, Precision in Equation~\ref{F1} shows in Equation~\ref{precision} and Recall in Equation~\ref{F1} shows in Equation~\ref{recall}.

     \item [$\bullet$]
     \textbf{KS} \cite{hodges1958significance}: The KS (Kolmogorov–Smirnov) value measures whether two independent distributions are similar or not. After generating cumulative probability, it searches for the maximum distance between two distributions. The smaller the distance, the more similar these two distributions are. In our case, the two distributions are predicted scores for $n$ links in the set of non-existent edges, and those for links in the test set, respectively.
     \begin{equation}
         KS = \sup _{u} | F_{m}(u)- G_{n}(u)|
     \end{equation}
     
     \item [$\bullet$]
     \textbf{NDCG@k (Normalized Discounted Cumulative Gain calculated by top $K$ links)}: NDCG@K cares about whether a link in the test set is placed in the front position of the rank list.
    \begin{equation}
    DCG@K =\sum_{i=1}^K \frac{2^{rel_i}-1}{\log_2(i+1)} \label{eq:dcg}
    \end{equation}
    \begin{equation}
    NDCG@K =\frac{DCG@K}{IDCG}  \label{eq:ndcgMetric}
    \end{equation}

where $rel_i$ means the relevance of the recommendation result of position i. IDCG represents a list of the best recommended results returned by a user of the recommendation system.
%that is, assuming that the returned results are sorted by relevance, the most relevant results are placed first and the DCG of this sequence is IDCG. 

     \item [$\bullet$]
     \textbf{MAP@K (Mean Average Precision by top-K links)} \cite{liang2016modeling}: MAP@K measures the mean of top-k items' average precision.
     \begin{equation}
         \text {Average Precision@k}=\sum_{n=1}^{k} \frac{\text { Precision }@n}{\min \left(n,\left|\mathbf{y}_{\text {test}}\right|\right)}
     \end{equation}
\end{itemize}

\begin{comment}
\begin{equation}
F1= \frac{2 \times Recall \times Precision}{Recall + Precision}
\label{F1}
\end{equation}

\begin{equation}
Precision = \frac{Number \ of \ correct \ predictions}{Number \ of \ total \ predictions}
\label{precision}
\end{equation}

\begin{equation}
Recall = \frac{Number \ of \ correct \ predictions}{Number \ of \ total \ test \ set}
\label{recall}
\end{equation}
\end{comment}
%%%%%%%%%%%%%%%%%%%%%%%%%%%%%%%%%%%%%%%%%%%%%%%%%%%%%%%%%%%%%%%%%%%%%%

\subsubsection{Parameter Setup}
We empirically adopt the optimal hyper-parameter settings. 
For the proposed LP-UIT method%\footnote{The source codes of LP-UIT can be found on: \url{github.com/Link-Prediction/LP-UIT}.}
, we apply two layers GCN in \emph{Node Representation}. The hidden sizes of GCN are $64$ and $16$ on Zhihu, $64$ and $64$   on Epinions. We use Adam optimizer with the initial learning rate $0.01$ on Zhihu and $5e-3$ on Epinions, respectively. 
The $L_2$ penalty is set to $5e-4$ on Zhihu and $5e-6$ on Epinions. Moreover, the dropout probability is $0.5$ on both Zhihu and Epinions datasets.
For deepMDBN method, we apply a three-layer MLP with hidden sizes $64$, $128$, $32$ on Zhihu and $256$, $64$, $32$ on Epinions, respectively. The initial learning rate is set to $0.01$ on Zhihu and $0.001$ on Epinions, respectively. 
For ARGA method, the hidden sizes are $128$, $64$ on Zhihu, and $256$, $128$ on Epinions. The dropout is set to $0.0$ on Zhihu and $0.2$ on Epinions, respectively. 
Noted that for LP-UIT and the best baseline, we run each experiment five times and conduct pair-wise t-test to validate the significance of the performance difference.

%%%%%%%%%%%%%%%%%%%%%%%%%%%%%%%%%%%%%%%%%%%%%%%%%%%%%%%%%%%%%%%%%%%%%%
\subsection{Experiment Results}

Here, we display the results of our evaluations to answer the aforementioned RQs, based on which we also provide corresponding explanations and discussions.

\subsubsection{Effectiveness of LP-UIT over Baseline Methods (RQ1)}
To demonstrate the effectiveness of LP-UIT, we compare it with other state-of-the-art baseline methods in terms of AUC, KS, NDCG@$500$ and MAP@$500$. The comparative results on the two datasets are present in Table \ref{tb:comparativeResults}. We have some interesting observations as below:
(1) the performance of LP-UIT is better than other baselines, validating the effectiveness of our framework;
(2) the deep learning-based methods (deepMDBN and ARGA) perform better than other classical methods, demonstrating the capability of deep learning techniques for link prediction; and
(3) our model outperforms deepMDBN, validating the effectiveness of graph convolutional networks. Furthermore, our model outperforms ARGA, indicating that learning word embedding and node embedding separately are more effective than learning user embedding directly (i.e., treating word embedding as the input of graph model). 

\begin{figure}[htbp]
    \centering
	\footnotesize
	\begin{tikzpicture}
      \matrix[
          matrix of nodes,
          draw,
          inner sep=0.1em,
          ampersand replacement=\&,
          font=\scriptsize,
          anchor=south
        ]
        { 
		\ref{plots:DEEP} deepMDBN
		\ref{plots:ARGA} ARGA
		\ref{plots:UIT} LP-UIT\\
          };
    \end{tikzpicture}\\
\begin{tikzpicture}
	\begin{groupplot}[group style={
		group name=myplot,
		group size= 2 by 2,  horizontal sep=1.5cm}, 
		height=4cm, width=4cm,
	ylabel style={yshift=-0.1cm},
	every tick label/.append style={font=\scriptsize}
	]
\nextgroupplot[
ylabel=NDCG@K (Zhihu),
xmin=200,
xmax=1000,
ymin=0.78,
ymax=1.01,
ytick={0.80, 0.9, 1},
ytick pos=left,
mark size=1.2pt
]
\addplot coordinates {
(200,0.8274)
(300, 0.8409)
(400, 0.8646)
(500, 0.8864)
(600, 0.8870)
(700, 0.8850)
(800, 0.8916)
(900, 0.8996)
(1000, 0.9017)

};\label{plots:DEEP}
\addplot coordinates {
(200,0.9109)
(300, 0.9109)
(400, 0.9147)
(500, 0.9155)
(600, 0.9108)
(700, 0.9037)
(800, 0.8972)
(900, 0.8947)
(1000, 0.8952)
};\label{plots:ARGA}
\addplot coordinates {
(200,0.9054)
(300, 0.9144)
(400, 0.9255)
(500, 0.9358)
(600, 0.9431)
(700, 0.9486)
(800, 0.9528)
(900, 0.9571)
(1000, 0.9549)
};\label{plots:UIT}

\nextgroupplot[
ylabel=MAP@K (Zhihu),
xmin=200,
xmax=1000,
ymin=0.78,
ymax=1.01,
ytick={0.80, 0.9, 1},
%Xtick={0.0001,0.001,0.005,0.01,0.02},
ytick pos=left,
mark size=1.2pt
]
\addplot coordinates {
(200,0.8942)
(300, 0.8619)
(400, 0.8558)
(500, 0.8597)
(600, 0.8652)
(700, 0.8680)
(800, 0.8706)
(900, 0.8738)
(1000, 0.8775)

};\label{plots:DEEP}
\addplot coordinates {
(200,0.9050)
(300, 0.9067)
(400, 0.9128)
(500, 0.9130)
(600, 0.9127)
(700, 0.9113)
(800, 0.9095)
(900, 0.9076)
(1000, 0.9059)
};\label{plots:ARGA}
\addplot coordinates {
(200,0.9330)
(300, 0.9205)
(400, 0.9189)
(500, 0.9212)
(600, 0.9246)
(700, 0.9280)
(800, 0.9314)
(900, 0.9346)
(1000, 0.9371)
};\label{plots:UIT}

%%%%%
\nextgroupplot[
ylabel=NDCG@K (Epinions),
xmin=200,
xmax=1000,
ymin=0.78,
ymax=1.01,
ytick={0.80, 0.9, 1},
ytick pos=left,
mark size=1.2pt
]
\addplot coordinates {
(200,0.9166)
(300, 0.9387)
(400, 0.9509)
(500, 0.9396)
(600, 0.9464)
(700, 0.9515)
(800, 0.9564)
(900, 0.9549)
(1000, 0.9545)

};\label{plots:DEEP}
\addplot coordinates {
(200,0.8329)
(300, 0.8612)
(400, 0.8850)
(500, 0.8922)
(600, 0.9014)
(700, 0.9047)
(800, 0.9124)
(900, 0.9194)
(1000, 0.9236)
};\label{plots:ARGA}
\addplot coordinates {
(200,0.9761)
(300, 0.9797)
(400, 0.9798)
(500, 0.9750)
(600, 0.9663)
(700, 0.9655)
(800, 0.9609)
(900, 0.9635)
(1000, 0.9624)
};\label{plots:UIT}

\nextgroupplot[
ylabel=MAP@K (Epinions),
xmin=200,
xmax=1000,
ymin=0.68,
ymax=1.01,
ytick={0.70, 0.8,0.9, 1},
%Xtick={0.0001,0.001,0.005,0.01,0.02},
ytick pos=left,
mark size=1.2pt
]
\addplot coordinates {
(200,0.9215)
(300, 0.9346)
(400, 0.9440)
(500, 0.9486)
(600, 0.9505)
(700, 0.9523)
(800, 0.9543)
(900, 0.9558)
(1000, 0.9568)

};\label{plots:DEEP}
\addplot coordinates {
(200,0.7860)
(300, 0.8236)
(400, 0.8477)
(500, 0.8629)
(600, 0.8741)
(700, 0.8820)
(800, 0.8889)
(900, 0.8948)
(1000, 0.8997)
};\label{plots:ARGA}
\addplot coordinates {
(200, 0.9764)
(300, 0.9765)
(400, 0.9778)
(500, 0.9772)
(600, 0.9760)
(700, 0.9739)
(800, 0.9722)
(900, 0.9708)
(1000, 0.9698)
};\label{plots:UIT}

\end{groupplot}

\end{tikzpicture}\vspace{-3mm}
   \caption{The performance of different K on two datasets.\vspace{-3mm}
    }\
\label{fig:kvalue}
\end{figure}
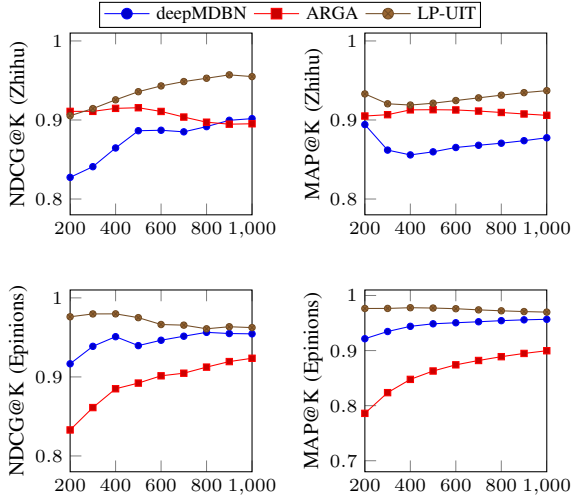

We also explore the performance of different approaches in terms of NDCG and MAP by varying $K\in\{200,300,400,500,600,700,800,900,1000\}$. Figure \ref{fig:kvalue} depicts the comparative results and shows that our model consistently performs better under almost all scenarios, further validating the superiority of LP-UIT over other DL-based methods (i.e., deepMDBNAND and ARGA).
%Figure \ref{fig:kvalue} reports the NDCG@K and MAP@K ($K=\{300,400,500,600,700,800,900,1000\}$) on two datasets by comparing LP-UIT with deep learning-based methods (deepMDBNAND and ARGA), which further proves the superiority of our model. 
%%%%%%%%%%

%%%%%%%%%%%%%%%%%%%%%%%%%%%%%%%%%%%%%%%%%%%%%%%%%%%%%%%%%%%%%%%%%%%%%%
\subsubsection{Ablation Study (RQ2)}
\begin{figure}[htb]
    \centering
	\footnotesize
	\begin{tikzpicture}
    %--define the legend by ourself
      \matrix[
          matrix of nodes,
          draw,
          inner sep=0.1em,
          ampersand replacement=\&,
          font=\scriptsize,
        ]
        { 
        \ref{plots:tplot1} No-long
		\ref{plots:tplot2} No-short
        \ref{plots:tplot3} No-link
		\ref{plots:tplot4} No-attention
		\ref{plots:tplot5} LP-UIT\\
          };
    \end{tikzpicture}\\
	
	\begin{tikzpicture}
	\begin{groupplot}[group style={
		group name=myplot,
		group size= 2 by 2,  horizontal sep=1.5cm}, 
		height=4cm, width=4cm,
	ylabel style={yshift=-0.1cm},
	%every tick/.append style={font=\tiny}
	every tick label/.append style={font=\scriptsize}
	]
	
	% AUC
	%\hspace{-0.2in}
	\nextgroupplot[ybar=0.10,
	bar width=0.55em,
	ylabel={AUC},
	scaled ticks=false,
	yticklabel style={/pgf/number format/.cd,fixed,precision=3},
	ymin=0.6, ymax=1.02,
	enlarge x limits=0.4,
	symbolic x coords={Zhihu,Epinions},
	ylabel style = {font=\scriptsize},
	xtick=data,
	ytick={0.6,0.8,1},
	]
	\addplot coordinates {
		(Zhihu,0.9208) (Epinions, 0.8885)};\label{plots:tplot1}
	\addplot coordinates {
		(Zhihu,0.9430) (Epinions, 0.9030)};\label{plots:tplot2}
	\addplot coordinates {
		(Zhihu,0.9410) (Epinions, 0.8970)};\label{plots:tplot3}
	\addplot coordinates {
		(Zhihu,0.9326) (Epinions, 0.8915)};\label{plots:tplot4}
	\addplot coordinates {
		(Zhihu,0.9516) (Epinions, 0.9048)};\label{plots:tplot5}
 
 	% KS
	%\hspace{-0.2in}
	\nextgroupplot[ybar=0.10,
	bar width=0.55em,
	ylabel={KS},
	scaled ticks=false,
	yticklabel style={/pgf/number format/.cd,fixed,precision=3},
	ymin=0.4, ymax=0.82,
	enlarge x limits=0.4,
	symbolic x coords={Zhihu,Epinions},
	ylabel style = {font=\scriptsize},
	xtick=data,
	ytick={0.4,0.6,0.8},
	]
	\addplot coordinates {
		(Zhihu,0.6777) (Epinions, 0.6019)};\label{plots:tplot1}
	\addplot coordinates {
		(Zhihu,0.7666) (Epinions, 0.6507)};\label{plots:tplot2}
	\addplot coordinates {
		(Zhihu,0.7513) (Epinions, 0.6492)};\label{plots:tplot3}
	\addplot coordinates {
		(Zhihu,0.7332) (Epinions, 0.6362)};\label{plots:tplot4}
	\addplot coordinates {
		(Zhihu,0.7815) (Epinions, 0.6594)};\label{plots:tplot5}
		
	% NDCG@500
	%\hspace{1in}
    \nextgroupplot[ybar=0.10,
	bar width=0.55em,
	ylabel={NDCG@500},
	scaled ticks=false,
	yticklabel style={/pgf/number format/.cd,fixed,precision=3},
	ymin=0.6, ymax=1.02,
	enlarge x limits=0.4,
	symbolic x coords={Zhihu,Epinions},
	ylabel style = {font=\scriptsize},
	xtick=data,
	ytick={0.6,0.8,1},
	]
	\addplot coordinates {
		(Zhihu,0.8960) (Epinions, 0.9354)};\label{plots:tplot1}
	\addplot coordinates {
		(Zhihu,0.9029) (Epinions, 0.9517)};\label{plots:tplot2}
	\addplot coordinates {
		(Zhihu,0.9137) (Epinions, 0.9356)};\label{plots:tplot3}
	\addplot coordinates {
		(Zhihu,0.8902) (Epinions, 0.9494)};\label{plots:tplot4}
	\addplot coordinates {
		(Zhihu,0.9385) (Epinions, 0.9750)};\label{plots:tplot5}
    
	% MAP@500
	%\hspace{-0.2in}
	\nextgroupplot[ybar=0.10,
	bar width=0.55em,
	ylabel={MAP@500},
	scaled ticks=false,
	yticklabel style={/pgf/number format/.cd,fixed,precision=3},
	ymin=0.6, ymax=1.02,
	enlarge x limits=0.4,
	symbolic x coords={Zhihu,Epinions},
	ylabel style = {font=\scriptsize},
	xtick=data,
	ytick={0.6,0.8,1},
	]
	\addplot coordinates {
		(Zhihu,0.9030) (Epinions, 0.9331)};\label{plots:tplot1}
	\addplot coordinates {
		(Zhihu,0.8691) (Epinions, 0.9562)};\label{plots:tplot2}
	\addplot coordinates {
		(Zhihu,0.8918) (Epinions, 0.9408)};\label{plots:tplot3}
	\addplot coordinates {
		(Zhihu,0.8714) (Epinions, 0.9444)};\label{plots:tplot4}
	\addplot coordinates {
		(Zhihu,0.9212) (Epinions, 0.9772)};\label{plots:tplot5}

	\end{groupplot}
    \end{tikzpicture}\vspace{-3mm}
    \caption{The performance of different variants.}\vspace{-3mm}
    \label{fig:variants}
\end{figure}
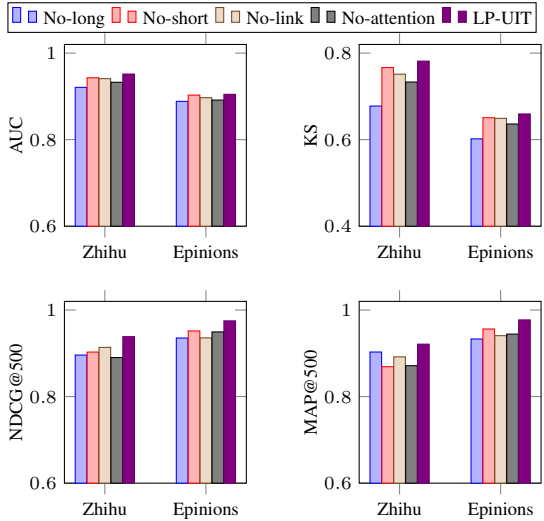

LP-UIT considers both user and topology information. To explore the effectiveness of each part, we compare LP-UIT  with four variants:
(1) ``No-long" ignores long-term interests;
(2) ``No-short" ignores short-term interests;
(3) ``No-link" removes weak links and user social influence information; and 
(4) ``No-attention" does not use the attention layer to get the importance from word to node. It simply concatenates $\mathbf{w}_u$ and $\mathbf{t}_u$ to generate the user embedding $\mathbf{x}_u$.
The comparative performance of LP-UIT and the four different variants are depicted in Figure \ref{fig:variants}.

As shown in Figure \ref{fig:variants}, LP-UIT performs better than the four variants, validating the effectiveness of our designs. Besides, No-short performs better than No-long, implying that only considering long-term interests is better than only considering short-term interests. This might be due to that people may prefer establishing links with similar long-term interests rather than with users who have similar short-term interests.
Moreover, the comparison between LP-UIT and No-link indicates that user social influence and weak links affect the performance of link prediction. This is because that people will be more willing to form links with celebrities or other users they have interacted with before. We have also experimentally verified the rationality of the attention mechanism of our model.

%%%%%%%%%%%%%%%%%%%%%%%%%%%%%%%%%%%%%%%%%%%%%%%%%%%%%%%%%%%%%%%%%%%%%%
\subsubsection{Sensitivity of Hyper-parameters (RQ3)}
\begin{figure}[htbp]
    \centering
    \includegraphics[width=7cm]{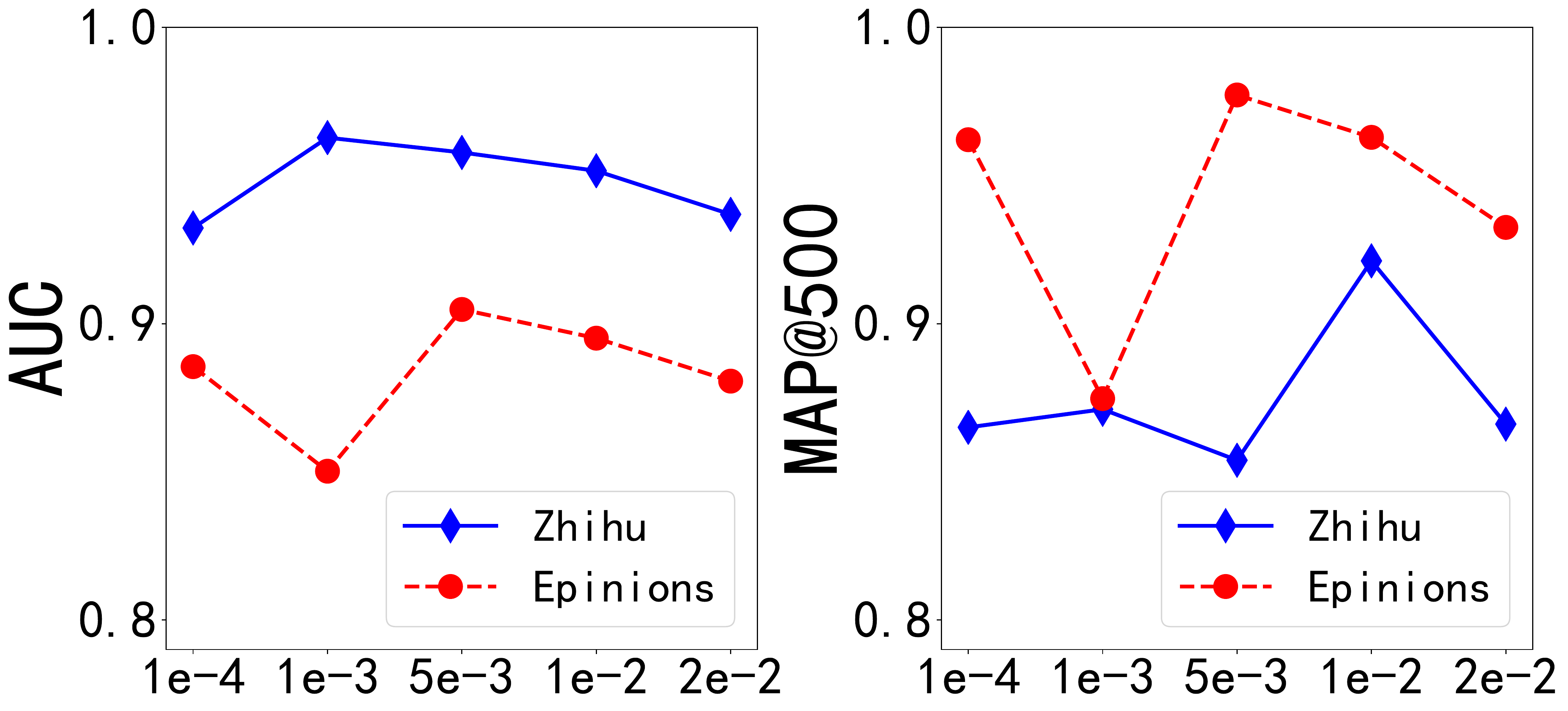}
    \caption{The performance of different learning rate.}
    \label{fig:learningrate}
\end{figure}

\begin{figure}[htbp]
    \centering
    \includegraphics[width=7cm]{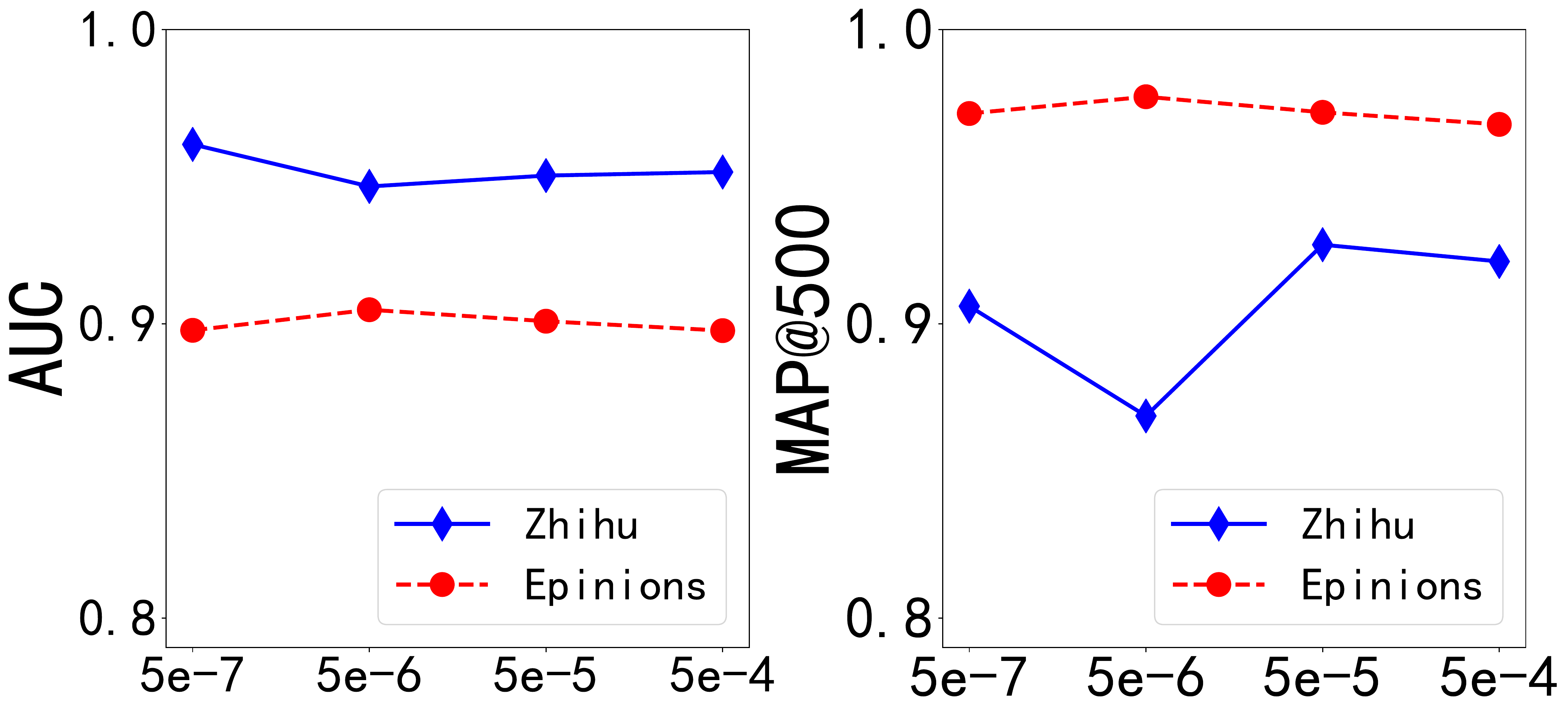}
    \caption{The performance of different $L_2$ penalty.}
    \label{fig:l2penalty}
\end{figure}

We investigate the impact of learning rate $lr$ and $L_2$ penalty on LP-UIT model, by deploying a grid search in the range of $\{0.0001, 0.001, 0.005, 0.01, 0.02\}$ and $\{5e-7, 5e-6, 5e-5, 5e-4\}$ for $lr$ and $L_2$ penalty, respectively. Figures \ref{fig:learningrate} and \ref{fig:l2penalty} show the experiment results.
Generally speaking, our method is comparatively insensitive to the two hyper-parameters. 
\vspace{-3mm}

%%%%%%%%%%%%%%%%%%%%%%%%%%%%%%%%%%%%%%%%%%%%%%%%%%%%%%%%%%%%%%%%%%%%%%
\section{Conclusions}
In this paper, towards the link prediction problem in information-seeking oriented social networks, we summarized three types of features which are derived from three information sources of three modalities. That is, users' long-term and short-term interests from textual information, users' social influence and weak links regarding interactions between each user pair from numerical information, and topological features from graph (network) information.
In this view, we proposed a novel multi-modal framework (called LP-UIT) which considered and fused a relative comprehensive set of features from different information sources. Comprehensive experiments on two real-world datasets showed that our model outperformed other state-of-the-art methods, and also validated the effectiveness of each component in LP-UIT.

For future study, we will consider to design more effective and advanced method to fuse different information sources for link prediction.

\section*{Acknowledgment}
This work was supported in part by the National Natural Science Foundation of China (Grant No. 71601104, 71601116, 71771141), the Shanghai Natural Science Foundation of China (Grant No. 21ZR1421900), and the Graduate Innovation Fund of Shanghai University of Finance and Economics (Grant No. CXJJ-2020-431).




%\bibliographystyle{IEEEtran}

\bibliography{myrefer}

\begin{thebibliography}{10}
\providecommand{\url}[1]{#1}
\csname url@samestyle\endcsname
\providecommand{\newblock}{\relax}
\providecommand{\bibinfo}[2]{#2}
\providecommand{\BIBentrySTDinterwordspacing}{\spaceskip=0pt\relax}
\providecommand{\BIBentryALTinterwordstretchfactor}{4}
\providecommand{\BIBentryALTinterwordspacing}{\spaceskip=\fontdimen2\font plus
\BIBentryALTinterwordstretchfactor\fontdimen3\font minus
  \fontdimen4\font\relax}
\providecommand{\BIBforeignlanguage}[2]{{%
\expandafter\ifx\csname l@#1\endcsname\relax
\typeout{** WARNING: IEEEtran.bst: No hyphenation pattern has been}%
\typeout{** loaded for the language `#1'. Using the pattern for}%
\typeout{** the default language instead.}%
\else
\language=\csname l@#1\endcsname
\fi
#2}}
\providecommand{\BIBdecl}{\relax}
\BIBdecl

\bibitem{he2017latent}
Z.~He, Z.~Cai, and J.~Yu, ``Latent-data privacy preserving with customized data
  utility for social network data,'' \emph{IEEE Transactions on Vehicular
  Technology}, vol.~67, no.~1, pp. 665--673, 2017.

\bibitem{katz1973uses}
E.~Katz, J.~G. Blumler, and M.~Gurevitch, ``Uses and gratifications research,''
  \emph{The Public Opinion Quarterly}, vol.~37, no.~4, pp. 509--523, 1973.

\bibitem{bae2018understanding}
M.~Bae, ``Understanding the effect of the discrepancy between sought and
  obtained gratification on social networking site users' satisfaction and
  continuance intention,'' \emph{Computers in Human Behavior}, vol.~79, pp.
  137--153, 2018.

\bibitem{yang2015network}
C.~Yang, Z.~Liu, D.~Zhao, M.~Sun, and E.~Y. Chang, ``Network representation
  learning with rich text information,'' in \emph{International Joint
  Conference on Artificial Intelligence}, vol. 2015, 2015, pp. 2111--2117.

\bibitem{liao2018attributed}
L.~Liao, X.~He, H.~Zhang, and T.-S. Chua, ``Attributed social network
  embedding,'' \emph{IEEE Transactions on Knowledge and Data Engineering},
  vol.~30, no.~12, pp. 2257--2270, 2018.

\bibitem{pan2018adversarially}
S.~Pan, R.~Hu, G.~Long, J.~Jiang, L.~Yao, and C.~Zhang, ``Adversarially
  regularized graph autoencoder for graph embedding,'' in \emph{International
  Joint Conference on Artificial Intelligence}, 2018.

\bibitem{simonovsky2018graphvae}
M.~Simonovsky and N.~Komodakis, ``Graphvae: Towards generation of small graphs
  using variational autoencoders,'' in \emph{International Conference on
  Artificial Neural Networks}.\hskip 1em plus 0.5em minus 0.4em\relax Springer,
  2018, pp. 412--422.

\bibitem{ma2018constrained}
T.~Ma, J.~Chen, and C.~Xiao, ``Constrained generation of semantically valid
  graphs via regularizing variational autoencoders,'' in \emph{Advances in
  Neural Information Processing Systems}, 2018, pp. 7113--7124.

\bibitem{de2018molgan}
N.~De~Cao and T.~Kipf, ``Molgan: An implicit generative model for small
  molecular graphs,'' \emph{ICML 2018 workshop on Theoretical Foundations and
  Applications of Deep Generative Models}, 2018.

\bibitem{mutlu2019review}
E.~C. Mutlu and T.~A. Oghaz, ``Review on graph feature learning and feature
  extraction techniques for link prediction,'' \emph{arXiv preprint
  arXiv:1901.03425}, 2019.

\bibitem{li2014modeling}
L.~Li, L.~Zheng, F.~Yang, and T.~Li, ``Modeling and broadening temporal user
  interest in personalized news recommendation,'' \emph{Expert Systems with
  Applications}, vol.~41, no.~7, pp. 3168--3177, 2014.

\bibitem{yin2014temporal}
H.~Yin, B.~Cui, L.~Chen, Z.~Hu, and Z.~Huang, ``A temporal context-aware model
  for user behavior modeling in social media systems,'' in \emph{International
  Conference on Management of Data}, 2014, pp. 1543--1554.

\bibitem{spasojevic2014lasta}
N.~Spasojevic, J.~Yan, A.~Rao, and P.~Bhattacharyya, ``Lasta: Large scale topic
  assignment on multiple social networks,'' in \emph{International Conference
  on Knowledge Discovery and Data Mining}, 2014, pp. 1809--1818.

\bibitem{jiang2015modeling}
B.~Jiang and Y.~Sha, ``Modeling temporal dynamics of user interests in online
  social networks,'' \emph{Procedia Computer Science}, vol.~51, pp. 503--512,
  2015.

\bibitem{piao2018inferring}
G.~Piao and J.~G. Breslin, ``Inferring user interests in microblogging social
  networks: a survey,'' \emph{User Modeling and User-Adapted Interaction},
  vol.~28, no.~3, pp. 277--329, 2018.

\bibitem{wellman1997electronic}
B.~Wellman, ``An electronic group is virtually a social network,''
  \emph{Culture of the Internet}, vol.~4, pp. 179--205, 1997.

\bibitem{zhang2011intrank}
L.~Zhang, H.~Fang, W.~K. Ng, and J.~Zhang, ``Intrank: Interaction ranking-based
  trustworthy friend recommendation,'' in \emph{International Conference on
  Trust, Security and Privacy in Computing and Communications}.\hskip 1em plus
  0.5em minus 0.4em\relax IEEE, 2011, pp. 266--273.

\bibitem{kelman1958compliance}
H.~C. Kelman, ``Compliance, identification, and internalization three processes
  of attitude change,'' \emph{Journal of Conflict Resolution}, vol.~2, no.~1,
  pp. 51--60, 1958.

\bibitem{marsden1993network}
P.~V. Marsden and N.~E. Friedkin, ``Network studies of social influence,''
  \emph{Sociological Methods \& Research}, vol.~22, no.~1, pp. 127--151, 1993.

\bibitem{huo2018link}
Z.~Huo, X.~Huang, and X.~Hu, ``Link prediction with personalized social
  influence,'' in \emph{AAAI Conference on Artificial Intelligence}, vol.~32,
  no.~1, 2018.

\bibitem{perozzi2014deepwalk}
B.~Perozzi, R.~Al-Rfou, and S.~Skiena, ``Deepwalk: Online learning of social
  representations,'' in \emph{International Conference on Knowledge Discovery
  and Data Mining}, 2014, pp. 701--710.

\bibitem{grover2016node2vec}
A.~Grover and J.~Leskovec, ``node2vec: Scalable feature learning for
  networks,'' in \emph{International Conference on Knowledge Discovery and Data
  Mining}, 2016, pp. 855--864.

\bibitem{tang2015line}
J.~Tang, M.~Qu, M.~Wang, M.~Zhang, J.~Yan, and Q.~Mei, ``Line: Large-scale
  information network embedding,'' in \emph{International World Wide Web
  Conferences}, 2015, pp. 1067--1077.

\bibitem{kipf2017semi}
T.~N. Kipf and M.~Welling, ``Semi-supervised classification with graph
  convolutional networks,'' in \emph{International Conference on Learning
  Representations}, 2017.

\bibitem{li2016gated}
Y.~Li, D.~Tarlow, M.~Brockschmidt, and R.~Zemel, ``Gated graph sequence neural
  networks,'' in \emph{International Conference on Learning Representations},
  2016.

\bibitem{ying2018hierarchical}
Z.~Ying, J.~You, C.~Morris, X.~Ren, W.~Hamilton, and J.~Leskovec,
  ``Hierarchical graph representation learning with differentiable pooling,''
  in \emph{Advances in Neural Information Processing Systems}, 2018, pp.
  4800--4810.

\bibitem{velickovic2018graph}
P.~Veli{\v{c}}kovi{\'{c}}, G.~Cucurull, A.~Casanova, A.~Romero, P.~Li{\`{o}},
  and Y.~Bengio, ``Graph attention networks,'' in \emph{International
  Conference on Learning Representations}, 2018.

\bibitem{cox2008multidimensional}
M.~A. Cox and T.~F. Cox, ``Multidimensional scaling,'' in \emph{Handbook of
  Data Visualization}.\hskip 1em plus 0.5em minus 0.4em\relax Springer, 2008,
  pp. 315--347.

\bibitem{roweis2000nonlinear}
S.~T. Roweis and L.~K. Saul, ``Nonlinear dimensionality reduction by locally
  linear embedding,'' \emph{Science}, vol. 290, no. 5500, pp. 2323--2326, 2000.

\bibitem{tenenbaum2000global}
J.~B. Tenenbaum, V.~De~Silva, and J.~C. Langford, ``A global geometric
  framework for nonlinear dimensionality reduction,'' \emph{Science}, vol. 290,
  no. 5500, pp. 2319--2323, 2000.

\bibitem{cao2016deep}
S.~Cao, W.~Lu, and Q.~Xu, ``Deep neural networks for learning graph
  representations,'' in \emph{AAAI Conference on Artificial Intelligence},
  vol.~30, no.~1, 2016.

\bibitem{wang2016structural}
D.~Wang, P.~Cui, and W.~Zhu, ``Structural deep network embedding,'' in
  \emph{International Conference on Knowledge Discovery and Data Mining}, 2016,
  pp. 1225--1234.

\bibitem{huang2018multimodal}
F.~Huang, X.~Zhang, C.~Li, Z.~Li, Y.~He, and Z.~Zhao, ``Multimodal network
  embedding via attention based multi-view variational autoencoder,'' in
  \emph{International Conference on Multimedia Retrieval}, 2018, pp. 108--116.

\bibitem{huang2019network}
F.~Huang, X.~Zhang, J.~Xu, C.~Li, and Z.~Li, ``Network embedding by fusing
  multimodal contents and links,'' \emph{Knowledge-Based Systems}, vol. 171,
  pp. 44--55, 2019.

\bibitem{mikolov2013efficient}
T.~Mikolov, K.~Chen, G.~Corrado, and J.~Dean, ``Efficient estimation of word
  representations in vector space,'' \emph{arXiv preprint arXiv:1301.3781},
  2013.

\bibitem{salton1975vector}
G.~Salton, A.~Wong, and C.-S. Yang, ``A vector space model for automatic
  indexing,'' \emph{Communications of the ACM}, vol.~18, no.~11, pp. 613--620,
  1975.

\bibitem{newman2001clustering}
M.~E. Newman, ``Clustering and preferential attachment in growing networks,''
  \emph{Physical Review E}, vol.~64, no.~2, p. 025102, 2001.

\bibitem{tong2006fast}
H.~Tong, C.~Faloutsos, and J.-Y. Pan, ``Fast random walk with restart and its
  applications,'' in \emph{International Conference on Data Mining}.\hskip 1em
  plus 0.5em minus 0.4em\relax IEEE, 2006, pp. 613--622.

\bibitem{liu2015multimodal}
F.~Liu, B.~Liu, C.~Sun, M.~Liu, and X.~Wang, ``Multimodal learning based
  approaches for link prediction in social networks,'' in \emph{Natural
  Language Processing and Chinese Computing}.\hskip 1em plus 0.5em minus
  0.4em\relax Springer, 2015, pp. 123--133.

\bibitem{hodges1958significance}
J.~L. Hodges, ``The significance probability of the smirnov two-sample test,''
  \emph{Arkiv f{\"o}r Matematik}, vol.~3, no.~5, pp. 469--486, 1958.

\bibitem{liang2016modeling}
D.~Liang, L.~Charlin, J.~McInerney, and D.~M. Blei, ``Modeling user exposure in
  recommendation,'' in \emph{International World Wide Web Conferences}, 2016,
  pp. 951--961.

\end{thebibliography}
\bibliographystyle{IEEEtran}

\end{document}